\documentclass[journal]{IEEEtran}

\usepackage{graphicx}
\usepackage{amsmath}
\usepackage{booktabs}
\usepackage{multirow}
\usepackage{hyperref}
\usepackage{cite}

\ifCLASSINFOpdf
\else
\fi
\begin{document}
\bstctlcite{BSTcontrol}
%
\title{Human Situational Awareness and Workload during Heterogeneous Autonomous Vehicle Hazard Monitoring: A Cross-Platform Study}
%
%
%

\author{Ziyue~Wang,
        Yang~Xing,~\IEEEmembership{Senior Member, IEEE},
        Argyrios Zolotas,~\IEEEmembership{Senior Member, IEEE}, Adolfo Perrusquia,~\IEEEmembership{Senior Member, IEEE}, Weisi~Guo, and Antonios Tsourdos
\thanks{The authors are with the Faculty of Engineering and Applied Sciences, Cranfield University, United Kingdom. MK43 0AL.}
\thanks{This work was supported in part by the Royal Society Research Grant RGS/R1/231037.}}

\maketitle

\begin{abstract}
As autonomous ground and aerial vehicles become increasingly capable, human operators are transitioning from direct control to supervisory monitoring roles. Maintaining Situational Awareness (SA) during rare but safety critical hazard events remains a fundamental challenge, particularly when operators must monitor heterogeneous autonomous platforms with different spatial and cognitive demands. This study develops a unified experimental framework for investigating human SA and workload during supervisory monitoring of autonomous ground and aerial vehicles under hazard conditions. Eighteen participants completed six within-subject conditions combining three interface modalities, namely, visual-only, haptic-only, and visual-haptic, with two platform types, ground vehicle and aerial drone. SA was measured using Situation Awareness Rating Technique (SART), while workload was assessed using the NASA Task Load Index (NASA-TLX). The key findings are: 1) Aerial monitoring produced significantly lower overall SA, higher situational demand, and greater workload than ground monitoring, driven by elevated Demand and reduced Understanding rather than diminished Supply. 2) Workload was consistently coupled with situational demand across all conditions, suggesting a stable SA--workload mechanism. 3) Both haptic-only and visual-haptic significantly improved overall SA relative to visual-only without increasing workload, demonstrating that interface design can enhance awareness within the existing workload envelope. These findings establish a cross-platform behavioural baseline for designing multimodal interfaces that enhance operator SA without elevating workload in heterogeneous autonomous vehicle systems.
\end{abstract}


%
\IEEEpeerreviewmaketitle

\section{Introduction}
%
%
%
%
\IEEEPARstart{S}{ituational} Awareness (SA) is a prerequisite for effective human supervision of autonomous systems \cite{endsley2017toward, endsley2024sa, jiang2023situation}. Endsley \cite{endsley2017toward} defined SA as spanning three levels, perception of environmental elements, comprehension of the current situation, and projection of future states, where degradation at any level delays hazard detection and weakens intervention \cite{endsley2025designing}. The demand for sustained SA is growing as autonomous platforms enter civilian infrastructure at scale \cite{yan2024uam, meir2025human, lee2023hci}, and maintaining adequate SA remains a recognised challenge in conditionally automated driving \cite{avetisyan2026towards}, motivating efforts to predict and monitor SA in real time during supervisory tasks \cite{zhou2022eye}. On the ground, autonomous vehicles equipped with supervised driving systems are entering urban traffic in increasing numbers \cite{lee2023hci}, yet automation displaces operators from active control and continuous monitoring, leaving them susceptible to out-of-the-loop performance decrements that degrade SA \cite{mutzenich2021updating}. This degradation is particularly consequential in hazardous conditions, where delayed perception directly increases collision risk \cite{endsley2025designing}. In the airspace, low-altitude operations and electric vertical take-off and landing aircraft are expanding rapidly \cite{yan2024uam, zhou2026heterogeneous}, and human operators remain necessary to maintain SA in scenarios that exceed current autonomous capabilities \cite{xue2026necessity}. These ground and aerial systems are increasingly converging within shared urban environments, forming a key component of future smart transportation and smart city ecosystems \cite{zhou2026heterogeneous}. As a result, operators are expected to supervise multiple heterogeneous autonomous platforms from both remote-control centres \cite{meir2025human}, and onboard environment \cite{wei2024autonomous}, creating SA challenges that differ fundamentally from traditional in-vehicle operation \cite{mutzenich2021updating, meir2025human}.\

Supervisory monitoring of autonomous systems also imposes cognitive workload, which reflects the processing resources consumed by the task \cite{hart1988development, wickens2021engineering}. In autonomous vehicle supervision, workload presents a distinctive challenge: prolonged routine monitoring induces attentional disengagement \cite{meir2025human}, yet infrequent hazardous events force an abrupt transition to high cognitive demand at the moment rapid intervention is required \cite{wickens2021engineering}. SA and workload are theorised to compete for shared attentional resources \cite{wickens2002situation,endsley2024sa}, and this prediction has been confirmed in supervisory domains where routine monitoring is punctuated by infrequent critical events \cite{li2021experimental, lim2021adaptive}. Factor-analytic evidence further confirms that the attentional dimensions of SA overlap with perceived workload, while the comprehension dimension loads on a separate factor \cite{braarud2021investigating, selcon1991workload}. An interface that improves SA at the cost of elevated workload may therefore offer no net operational benefit \cite{wickens2021engineering}, motivating the concurrent analysis of both constructs rather than either in isolation.\

Despite the theoretical and practical importance of concurrent SA and workload evaluation, these two constructs have not been sufficiently analysed together across ground and aerial platforms under hazard conditions. It remains unknown whether the cross-platform differences in spatial complexity produce distinct patterns of SA degradation and workload elevation, or whether the SA--workload coupling observed in single-platform studies \cite{li2021experimental, lim2021adaptive} holds across heterogeneous environments. At the interface level, non-visual sensory channels have demonstrated potential for enhancing SA during hazardous events. Haptic feedback has shown promise for enhancing hazard detection in both ground and aerial domains \cite{wang2025cross, xing2026augmenting}, yet the haptic channel has been tested almost exclusively alongside visual information \cite{wang2025cross, xing2026augmenting}, and whether it can support SA as a standalone modality without elevating workload is not known.\

\looseness=-1
To address these gaps, the present study develops a unified experimental framework that measures SA and workload concurrently through the Situation Awareness Rating Technique (SART) \cite{taylor2017sart} and the NASA Task Load Index (NASA-TLX) \cite{hart1988development}. Using a within-subject experiment with 18 participants, SA and workload were examined across ground vehicle and aerial vehicle hazard scenarios under three information-presentation conditions: visual-only, representing conventional supervisory displays; haptic-only, designed to isolate the independent contribution of tactile information; and visual-haptic, representing combined visual and haptic information presentation. This design enables a systematic investigation of how SA and workload interact across different platforms and interface conditions during safety-critical hazard monitoring. This study contributes to the literature in three respects:\

\begin{itemize}
    \item A unified cross-platform hazard-monitoring experimental framework is developed to investigate human SA and workload during supervision of both ground and aerial autonomous vehicles, enabling direct comparison across environments with different spatial and cognitive demands.\
    \item SA--workload coupling is demonstrated to remain stable across heterogeneous platform types, while component-level analysis reveals that cross-platform SA degradation is driven by elevated situational Demand and reduced Understanding rather than diminished Supply, establishing a behavioural baseline for platform-specific interface design in heterogeneous autonomous systems.\
    \item The independent and combined effects of haptic and visual information on SA and workload are disentangled by including a haptic-only condition alongside visual-only and visual-haptic conditions, demonstrating that the haptic channel alone can improve SA without increasing workload.\
\end{itemize}

\looseness=-1
The remainder of this paper is organised as follows. Section 2 reviews related work on the joint analysis of SA and workload and on multimodal interface design for SA augmentation. Section 3 describes the experimental methodology, including the simulation environment, interface design, and statistical approach. Section 4 presents the results. Section 5 discusses the findings and their limitations, and Section 6 concludes the paper.\


\hfill 
 
\hfill

\section{Related Work}

\subsection{Joint Analysis and Measuring of SA and Cognitive Workload in Interface Evaluation}
How different information-presentation strategies affect SA and workload remains a central question in interface evaluation \cite{tan2020methodology}, with evidence demonstrating that the haptic channel can carry several distinct types of operationally relevant information \cite{chai2022when}. In the ground vehicle domain, wearable vibrotactile vests have conveyed obstacle location \cite{samsel2025comparison} and hazard notification \cite{fang2025vibrotactile}, directional haptic warnings have reduced brake response times for drivers with impaired vision \cite{xu2025directional}, and seat-based cues have shortened takeover reaction times \cite{hong2022effect} with static patterns outperforming dynamic patterns \cite{petermeijer2017comparing}. Two-stage warning systems improved SA and reduced driving stress compared with single-stage alerts \cite{ma2021takeover}, and combining visual, auditory, and haptic modalities produced superior takeover performance compared with any single channel \cite{yun2020multimodal}. Body location and vibration pattern significantly affect recognition accuracy \cite{samsel2025comparison, lisboa2024haptic}, and operators prefer simple, spatially direct configurations \cite{lee2024participatory}.\

SA and workload measurement in interface evaluation usually follows two established approaches \cite{taylor2017sart, selcon1991workload}. Endsley \cite{endsley1995measurement} formalised the distinction between objective and subjective methods.  Situation Awareness Global Assessment Technique (SAGAT) provides freeze-probe queries during task execution \cite{endsley2017toward} but is impractical for brief trial durations where freezing would disrupt the task. SART provides post-trial self-report ratings without interrupting task performance \cite{taylor2017sart}; a direct comparison found that SAGAT captured objective knowledge accuracy, while SART was more closely associated with operator confidence and self-assessed performance \cite{endsley1998comparative}. A meta-analysis of 678 effects from 77 papers reinforces the importance of examining SA components individually rather than relying solely on composite scores \cite{bakdash2022validity}.\

Evidence for dimensional overlap between SA and workload measures comes from multiple sources. A comparison of SART and NASA-TLX in a flight simulation suggested that the attentional dimensions of SART may partly capture workload rather than SA \cite{selcon1991workload}, and factor analysis of ratings from licensed control room operators confirmed that SA Demand and Supply loaded on the same factor as NASA-TLX mental demand, temporal demand, and effort, while Understanding loaded on a separate factor \cite{braarud2021investigating}. Both meta-analytic \cite{bakdash2022validity} and measurement-theoretic \cite{bolton2021level} evidence further confirm that the three SA dimensions operate on separate measurement scales, questioning the validity of combining them into a single composite score. Recent factor-analytic work using eye-tracking and vehicle kinetic data has further confirmed that SA decomposes into distinct cognitive and action-strategy factors rather than operating as a unitary construct \cite{lai2025framework}. These findings collectively point toward analysing the three SA constructs individually and pairing SA with an independent workload instrument.\

Despite the theoretical and empirical evidence for SA-workload coupling, two measurement limitations persist in current interface evaluation research. First, component-level SA analysis remains uncommon, with reviews noting that most studies report only aggregate scores \cite{vandemerwe2024agent}, and investigations measuring both SA and workload have relied on the composite SA index whose validity has been questioned \cite{selcon1991workload,bolton2021level}; in haptic interface research specifically, SA has been reported at the aggregate level \cite{wang2025cross, zhang2023physiological} without decomposition into the three SART constructs. Although peripheral haptic information has been shown to enhance SA without reducing workload in fully automated driving \cite{yusof2022reading}, those findings were obtained on a single ground platform without component-level SA decomposition or a standalone haptic condition. Second, SA and workload measurement remain separated in haptic interface studies. investigations have measured either workload without validated SA instruments \cite{huang2025aerohaptix, bodsworth2024seeing} or SA without workload \cite{wang2025cross, xing2026augmenting}, and no study has measured both constructs concurrently across multiple modalities and platforms. These limitations indicate a need for SA analysis paired with workload measurement in cross-platform supervisory monitoring research.\

\subsection{SA Augmentation through Multimodal Interfaces}
How different information-presentation strategies affect SA and workload remains a central question in interface evaluation. The multisensory information presentation has been studied across ground vehicle and aerial vehicle domains \cite{tan2020methodology}, and the resulting evidence demonstrates that, except the commonly used visual and audio interface, the haptic channel can carry several distinct types of operationally relevant information \cite{chai2022when}. In the ground vehicle domain, wearable vibrotactile vests have been used to convey obstacle location in road traffic \cite{samsel2025comparison}, and vibrotactile safety vests have been evaluated for real-time hazard notification in construction environments \cite{fang2025vibrotactile}. Directional haptic warnings have improved hazard detection for drivers with impaired vision, significantly reducing brake response times and collision rates \cite{xu2025directional}. Haptic cues have also demonstrated value for time-critical responses. Seat-based vibrotactile cues shortened takeover reaction times in conditionally automated driving \cite{hong2022effect}, with spatially static vibration patterns producing faster responses than dynamic sweeping patterns \cite{petermeijer2017comparing}, and two-stage warning systems have been shown to improve SA and reduce driving stress during takeover compared with single-stage alerts \cite{ma2021takeover}. Combining visual, auditory, and haptic warning modalities produced superior takeover performance compared with any single channel \cite{yun2020multimodal}. Across these applications, the design parameters of haptic interfaces have been systematically investigated. Body location and vibration pattern significantly affect recognition speed and interpretation accuracy \cite{samsel2025comparison, lisboa2024haptic}, and participatory design studies have confirmed that operators consistently prefer simple, spatially direct configurations over complex ones \cite{lee2024participatory}.\

In the aerial vehicle domain, wearable vibrotactile systems have conveyed obstacle information during teleoperation \cite{huang2025aerohaptix}, guided micro-quadrotor swarms through tactile cues \cite{tsykunov2019swarmtouch}, communicated flight envelope boundaries \cite{vanbaelen2021flying}, and improved teleoperation under limited visibility through hand-worn haptic-gesture interfaces \cite{macchini2020handworn}. Skin-interfaced systems have combined gesture control with tactile feedback to enhance environmental awareness during flight \cite{yiu2025skin}. These findings establish that the haptic channel possesses substantial information-carrying capacity across spatial \cite{spence2008tactile, chai2022when}, directional \cite{meng2015tactile, tan2020methodology}, and temporal \cite{meng2015tactile, tan2020methodology} domains. In heterogeneous platform operations \cite{zhou2026heterogeneous, meir2025human}, human-robot teams outperformed fully autonomous systems in search and rescue, with operator SA proving most critical when conditions exceeded autonomous capabilities \cite{endsley2025designing, robinson2024human}.\

\looseness=-1
Wickens \cite{wickens2008multiple, wickens2024misconceptions} proposed that tasks drawing on different perceptual modalities interfere less than tasks competing within the same channel. Two meta-analyses support this: adding a tactile channel improved performance with the greatest gains under high workload \cite{burke2006comparing}, and haptic feedback outperformed visual feedback for takeover requests during visually demanding tasks \cite{lan2026automated}. Similarly, acoustic cues have been shown to significantly improve SA in simulated accident situations, providing further evidence that non-visual channels enhance hazard awareness \cite{ju2022acoustic}. Visual-auditory-haptic warnings produced the best takeover performance among seven modality combinations \cite{yun2020multimodal}, though seat-based haptic cues introduced trade-offs between reaction time and steering stability \cite{hong2022effect}. Multimodal explanations have improved driver understanding during takeover \cite{lee2023investigating}, and visual-haptic feedback has enhanced perceptual alignment in aerial supervision \cite{wu2025enhancing}. Across both domains, haptic feedback has been examined primarily as a supplement to visual or auditory information \cite{wang2025cross, wu2025enhancing, yun2020multimodal, xing2026augmenting}, and theory predicts that separating channels should reduce cross-modal interference \cite{wickens2008multiple, wickens2024misconceptions, chai2022when}.\

Despite the above studies, two limitations constrain the current understanding of multimodal interfaces for SA augmentation. The haptic channel has been evaluated almost exclusively alongside visual or auditory information \cite{wang2025cross, wu2025enhancing, yun2020multimodal, xing2026augmenting}, and whether it can support SA as a standalone modality remains untested. Furthermore, cross-platform evaluations of haptic interfaces have not measured SA and workload concurrently \cite{wang2025cross, xing2026augmenting}, making it unclear whether observed SA gains come at a cognitive cost. To address these gaps, the present study measures SA and workload concurrently across ground and aerial vehicle platforms under three different interface conditions: visual-only, haptic-only, and visual-haptic.\

\section{Methodology}
\begin{figure*}
    \centering
    \includegraphics[width=\linewidth]{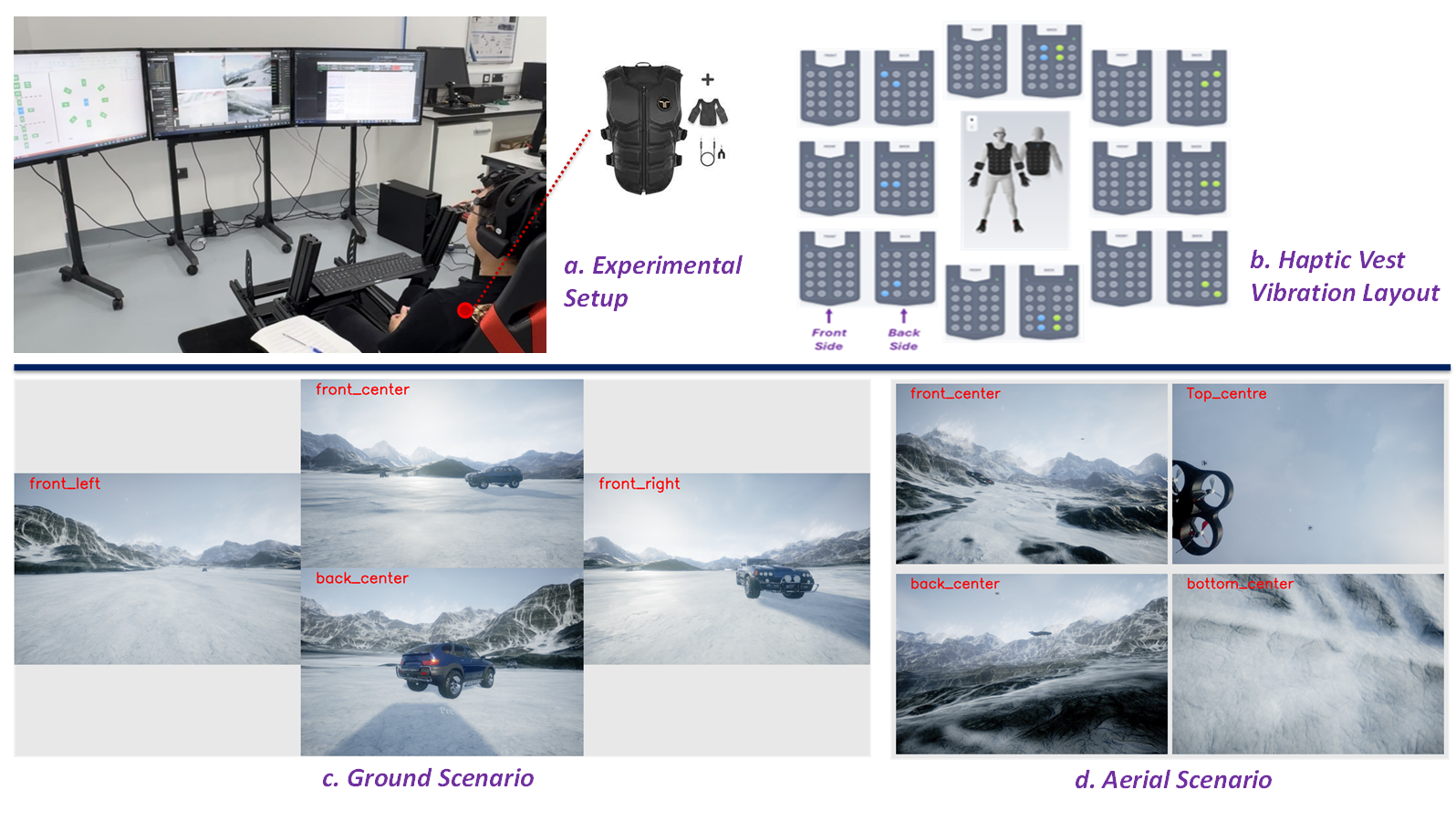}
    \caption{Experimental setup and scenario views. (a) The multi-display workstation and X40 bHaptics TactSuit vest worn by participants during the experiment. (b) Haptic vest vibration layout. Front and back panels encode hazard position jointly for three-dimensional spatial localisation. (c) Camera perspectives in the ground scenario, showing front-left, front-centre, front-right, and back-centre views. (d) Camera perspectives in the aerial scenario, showing front-centre, top -centre, back-centre, and bottom-centre views.}
    \label{fig:flowchart}
\end{figure*}

To investigate SA and workload during safety-critical monitoring of heterogeneous autonomous systems, a controlled laboratory experiment was developed using representative ground-vehicle and aerial-vehicle hazard scenarios. As illustrated in Fig.~\ref{fig:flowchart}, the experimental framework consisted of a high-fidelity simulation environment, a wearable vibrotactile interface, and a multi-display visual monitoring station. Ground and aerial scenarios were designed to represent two operational environments with distinct spatial characteristics, while visual-only, haptic-only, and visual-haptic conditions enabled systematic examination of different information-presentation strategies. Following each trial, participants reported their perception of the hazard events and completed subjective assessments of SA and workload. \

\begin{figure}
    \centering
    \includegraphics[width=\linewidth]{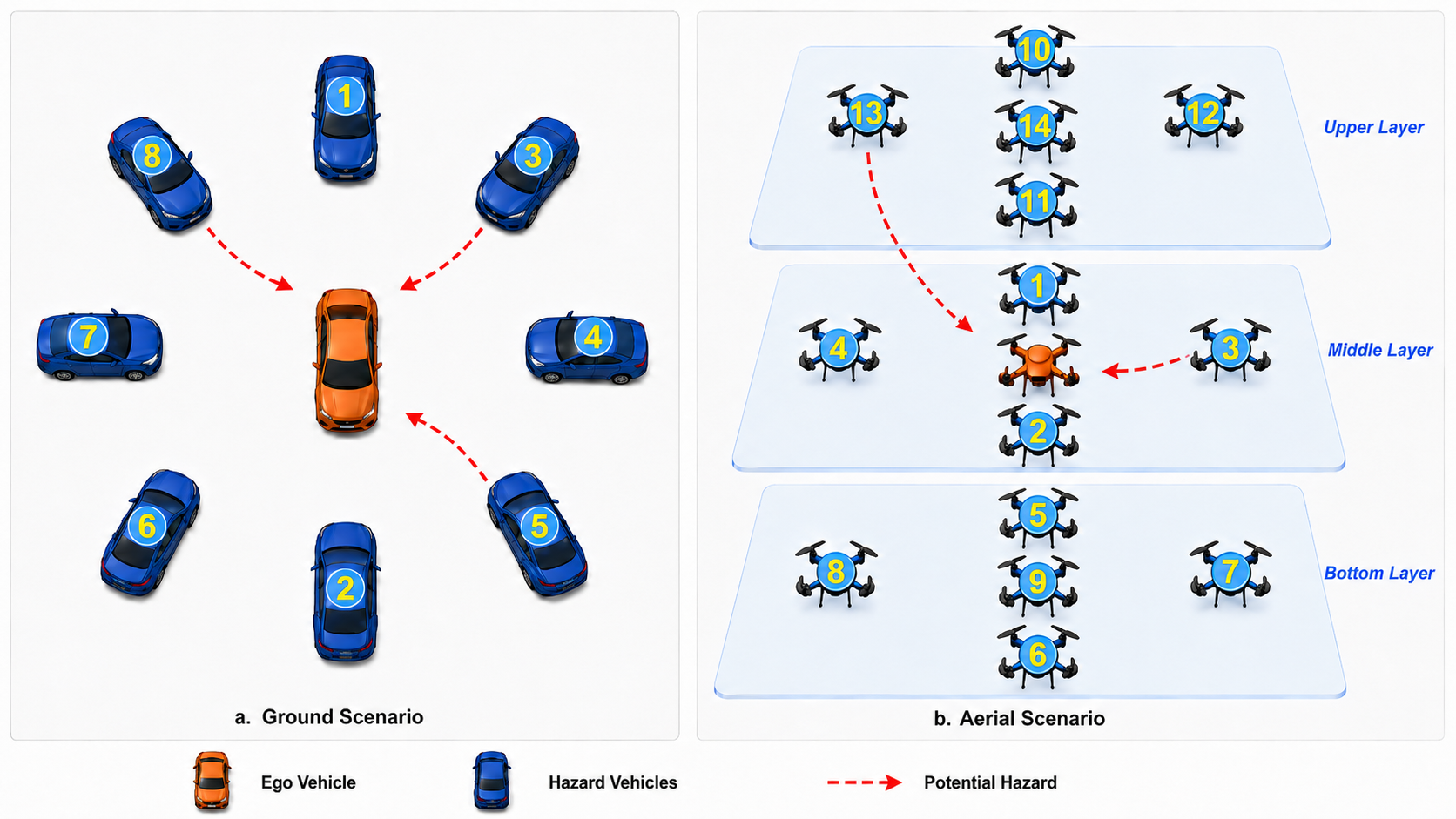}
    \caption{Spatial layout of hazard positions. (a) Ground scenario with eight surrounding vehicles distributed around the ego vehicle. (b) Aerial scenario with 14 surrounding drones distributed across three altitude layers.}
    \label{fig:position}
\end{figure}

The experiment simulates hazard monitoring in two autonomous vehicle scenarios with different spatial complexity. As shown in Fig.~\ref{fig:position}, there are 8 potential hazard positions in the ground scenario, distributed around the ego vehicle on a single horizontal plane, and 14 in the aerial scenario, distributed across three altitude layers.\

\subsection{Participants}
Eighteen participants, nine female and nine male, were recruited from Cranfield University between June and August 2025. Ages ranged from 24 to 35 years (M = 28.7,SD = 3.4). All were postgraduate students or research staff who reported normal or corrected-to-normal vision and no conditions affecting tactile perception. Four participants had prior exposure to the AirSim simulation environment. Because the experimental task required observation and reporting rather than manual vehicle control, prior simulation experience was not expected to influence performance, and these participants were retained. No additional screening criteria were applied. All participants volunteered without monetary compensation.\

Participants received an information sheet at least 24 hours before the session and signed a written consent form prior to data collection. The consent procedure included a 28-day withdrawal period during which participants could request removal of their data. No participant withdrew or was excluded from the analysis. Each participant completed all six conditions within a single session lasting approximately 1.5 to 2 hours. Ethical approval was obtained from the Cranfield University Research Ethics Committee (Ref. CURES/24590/2025).\

\subsection{Experimental Design}
The study employed a fully within-subject design with two independent variables: interface modality at three levels and scenario type at two levels. The three interface modalities were visual-only, haptic-only, and visual-haptic. In the visual-only condition, participants received environmental information solely through visual displays. In the haptic-only condition, the visual displays were deactivated and participants relied entirely on vibrotactile cues from a wearable vest. In the visual-haptic condition, both channels were available simultaneously. The two scenario types were ground vehicle and aerial drone. Each participant completed all six conditions with three repeated trials per condition, yielding 18 trials in total. The dependent variables were overall SA score and its three constituent components, Demand, Supply, and Understanding \cite{taylor2017sart}, and overall perceived workload from the NASA-TLX \cite{hart1988development}. SA was analysed at the component level because the three modalities were expected to affect different aspects of awareness, and aggregating to a single score would obscure these distinctions. The order of modality blocks was counterbalanced across participants so that each modality appeared first for one third of the sample. Within each interface modality block, scenario order was not systematically counterbalanced but varied across participants as a result of scheduling practicalities, producing a partial mix of ground-first and aerial-first orderings across the sample. The independent and dependent variables of the experiment are shown in Table~\ref{tab:variables}.\

\begin{table*}[t]
\caption{Experimental variables.}
\label{tab:variables}
\centering
\begin{tabular}{lll}
\toprule
\textbf{Variable Type} & \textbf{Variable} & \textbf{Levels / Instrument} \\
\multirow{2}{*}{Independent}
 & Interface modality & Visual-only, Haptic-only, Visual-haptic \\
 & Scenario type & Ground vehicle, Aerial drone \\
\midrule
\multirow{5}{*}{Dependent} 
 & Overall SA & SA composite score \\
 & Situational Demand & SA subscale \\
 & Situational Supply & SA subscale \\
 & Situational Understanding & SA subscale \\
 & Perceived workload & NASA-TLX weighted score \\
\bottomrule
\end{tabular}
\end{table*}

\subsection{Simulation Environment}
\looseness=-1
The experimental platforms for both ground and aerial scenarios were developed using Microsoft AirSim and Unreal Engine 4.27. During the experiment, participants faced three 48-inch monitors displaying the simulation environment. As shown in Fig.2, in the ground scenario, eight surrounding vehicles were positioned at fixed locations approximately 50 metres from the ego vehicle, distributed across front, rear, lateral, and diagonal directions to simulate a dense urban traffic environment. The ego vehicle drove autonomously for approximately ten seconds before entering the hazard zone, defined as the point at which it reached the nearest surrounding vehicle. Up to three vehicles then converged on the ego platform at 15 m/s. Vehicles were initiated sequentially by the programme with fractions of a second between initiations, so that the hazard events appeared nearly simultaneous to participants. Participants monitored the environment through four camera feeds providing front-centre, front-left, front-right, and back-centre perspectives, each with a 120-degree field of view.\ 

In the aerial scenario, 14 surrounding drones were positioned approximately 50 metres from the ego drone across three altitude layers: five in the upper layer, five in the middle layer, and four in the lower layer, simulating a congested low-altitude airspace. The ego drone flew autonomously for approximately ten seconds before entering the hazard zone, then hovered at the centre position while up to three drones converged at 15 m/s. The camera configuration replaced left and right views with top-centre and bottom-centre views to convey vertical spatial information. The aerial scenario was therefore more spatially complex than the ground scenario in two respects: hazards could originate from a larger number of positions, and participants needed to process vertical in addition to horizontal spatial information.\

\subsection{Interface Modalities}
Haptic feedback was delivered through a bHaptics TactSuit X40 vest equipped with 40 vibrotactile actuators, 20 on the front panel and 20 on the back panel. Haptic patterns consisted of simple dot vibrations at specific actuator locations, designed using the bHaptics Designer software. This design was informed by pilot testing, in which continuous sweeping vibrations proved effective for isolated hazard events but caused confusion under multi-hazard conditions; simple dot patterns provided clearer spatial discrimination. Each haptic cue lasted one second, and in trials with multiple converging hazards, participants experienced up to three sequential vibrations.\

For the ground scenario, only the back panel was active. Eight distinct haptic patterns were mapped to the eight surrounding vehicle positions, with each vibration location corresponding to the horizontal bearing of the approaching vehicle. For the aerial scenario, both panels were used: the back panel encoded horizontal bearing as in the ground condition, while the front panel conveyed altitude information, with upper-layer drones triggering vibration in the upper chest area and lower-layer drones triggering vibration in the abdominal area. Middle-layer drones activated only the back panel. This mapping decoupled vertical from horizontal spatial channels, though it required cognitive translation between vest location and spatial direction.\

\subsection{Experimental Procedure and Data Collection}
\begin{figure*}
    \centering
    \includegraphics[width=\textwidth]{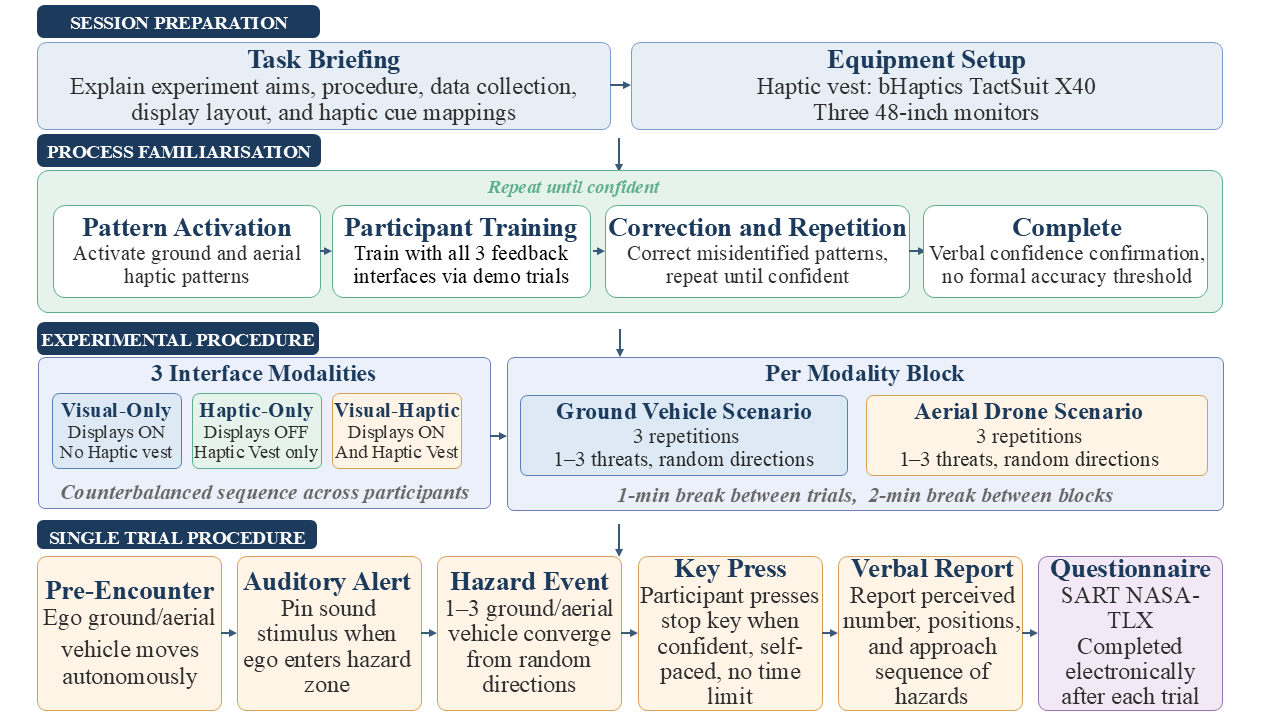}
    \caption{Experimental workflow. The session began with task briefing and equipment setup, followed by a familiarisation phase in which participants practised identifying haptic patterns until confident. The main procedure consisted of three counterbalanced modality blocks, each containing ground vehicle and aerial drone scenarios with three repetitions. Each trial followed a six-step sequence from the pre-encounter phase through to questionnaire completion.}
    \label{fig:procedure}
\end{figure*}

\looseness=-1
Fig.~\ref{fig:procedure} illustrates the experimental workflow, which comprised four stages: session preparation, process familiarisation, the main experimental procedure, and the single trial procedure.\

At the start of the session, the experimenter provided a detailed introduction covering the experimental task, the visual display layout for both ground and aerial scenarios, and the haptic cue mappings for each scenario. Participants then completed a haptic familiarisation phase in which the experimenter iteratively activated haptic patterns corresponding to ground and aerial hazard directions while participants tracked the stimuli and reported any discrepancies between their perceived vibration location and the intended direction displayed on screen. The experimenter corrected misidentifications and repeated patterns until each participant could consistently identify randomly generated single patterns. This phase typically required 8 to 10 minutes and ended when the participant verbally confirmed confidence in distinguishing the patterns; no formal accuracy criterion was imposed.\

Three timestamps were recorded per trial: programme start time, auditory alert time marking the onset of the encounter phase, and the participant's button press time indicating readiness to report. Each trial began with a pre-encounter phase of approximately ten seconds during which the ego vehicle or drone moved autonomously. When the ego platform entered the hazard zone, a one-second auditory alert signalled the onset of the hazard event. Up to three surrounding vehicles or drones then converged on the ego platform from randomised directions, with haptic cues triggered in sequence in the haptic-only and visual-haptic conditions. Participants faced no time constraint for their response; they pressed a keyboard key when confident in their assessment, then verbally reported the perceived number, positions, and sequence of the converging hazards. SART and NASA-TLX questionnaires were completed immediately after each trial. One-minute breaks were scheduled between trial sub-blocks and two-minute breaks between modality blocks, and participants could request additional pauses at any time. No participant reported fatigue affecting engagement or performance. The entire session lasted between 1.5 and 2 hours including equipment preparation, familiarisation, 18 formal trials, questionnaire completion, and rest breaks.\

\subsection{Measures and Statistical Analysis}
SA was assessed using the SART after each trial \cite{taylor2017sart}. Participants completed ten items after each trial, organised into three constructs: Demand, representing the attentional resources required by the task; Supply, representing the attentional resources available to the operator; and Understanding, representing the degree to which the operator comprehended the situation. Demand comprised three items: Instability, Complexity, and Variability of Situation. Supply comprised four items: Arousal, Concentration of Attention, Division of Attention, and Spare Mental Capacity. Understanding comprised three items: Information Quantity, Information Quality, and Familiarity with Situation. Each item was scored on a ten-point Likert scale. An overall SA score was computed using the standard overall SA formula:

\begin{equation}
\textit{Overall SA} = \sum_{i=1}^{3} U_i + \sum_{i=1}^{3} S_i - \sum_{i=1}^{3} D_i
\label{eq:overall_sa}
\end{equation}

\noindent where $U_i$, $S_i$, and $D_i$ denote the ratings for the $i$-th dimension of Understanding, Supply, and Demand, respectively.

Perceived workload was assessed using the NASA-TLX \cite{hart1988development}, with an overall score computed using the weighted procedure described by Hart \cite{hart2006nasa}.\

Prior to the primary analysis, all dependent variables were assessed for normality using the Shapiro-Wilk test. ANOVA was retained where departures were observed, given its robustness in balanced within-subject designs, with Greenhouse--Geisser correction applied where sphericity was violated. The primary analysis consisted of a two-way repeated-measures ANOVA for each dependent variable, with interface modality and scenario type as within-subject factors. Where the modality main effect reached significance, pairwise comparisons were conducted with Benjamini-Hochberg correction to control the false discovery rate \cite{benjamini1995controlling}. The scenario factor had only two levels and required no post-hoc testing. Effect sizes are reported as partial eta-squared for ANOVA effects and Hedges' g for pairwise contrasts. Effect sizes for pairwise contrasts are interpreted following Cohen's conventions, with $g = 0.2$ as small, 0.5 as medium, and 0.8 as large. Associations between workload and SA Demand were assessed using Spearman rank-order correlations to examine whether perceived workload and situational demand co-varied across conditions. Although factor-analytic evidence has shown that both situational dimensions of SA overlap with perceived workload \cite{braarud2021investigating, selcon1991workload}, SA Demand was selected for this analysis because it represents the task-imposed attentional cost, which most directly corresponds to the demand-on-capacity mechanism underlying perceived workload \cite{wickens2021engineering}. Post-hoc achieved power was computed for all effects. A significance threshold of $\alpha$ = 0.05 was adopted throughout.\

\section{Results}
This section examines how platform type and information-presentation strategy influenced SA and perceived workload during hazard monitoring. The analysis proceeds in three stages. First, overall SA is analysed to determine whether awareness differs across experimental conditions. Second, the three SART components, i.e., Demand, Supply, and Understanding, are examined individually to identify which aspects of SA contribute to any observed differences. Finally, perceived workload and its relationship with situational demand are investigated to explore the coupling between cognitive workload and SA. Descriptive statistics for all dependent variables are reported in Table~\ref{tab:descriptives}, while the results of the repeated-measures ANOVA are summarised in Table~\ref{tab:anova}.\

\subsection{Overall SA}
Table~\ref{tab:descriptives} presents the descriptive statistics for all dependent variables across the six experimental conditions. Overall SA scores were consistently higher in the ground scenario than in the aerial scenario for all three modalities. In the ground scenario, mean overall SA increased from 21.28 (SD = 6.75) under visual-only to 23.69 (SD = 6.19) under haptic-only and 23.69 (SD = 6.19)under visual-haptic. In the aerial scenario, modality differences were more pronounced: visual-only yielded a mean of 15.98 (SD = 5.17), compared with 21.52 (SD = 5.80) under haptic-only and 21.52 (SD = 8.85) under visual-haptic. The visual-only condition showed the largest difference between scenarios, with a mean overall SA drop of 5.30 from ground to aerial, while haptic-only and visual-haptic retained means above 21 in both scenarios.\

\begin{table*}[t]
\caption{Descriptive statistics for all dependent variables. Values are M $\pm$ SD, $N$ = 18. Bold font indicates the largest value in each condition.}
\label{tab:descriptives}
\centering
\begin{tabular}{llccccc}
\toprule
\textbf{Scenario} & \textbf{Interface} & \textbf{Overall SA} & \textbf{Demand} & \textbf{Supply} & \textbf{Understanding} & \textbf{Workload} \\
\midrule
\multirow{3}{*}{Ground}
 & Visual-only    & 21.28 $\pm$ 6.75          & \textbf{3.90 $\pm$ 1.46} & 4.50 $\pm$ 1.20          & 4.99 $\pm$ 0.91          & 4.74 $\pm$ 1.64 \\
 & Haptic-only    & 23.69 $\pm$ 6.19          & 3.46 $\pm$ 1.36          & 4.56 $\pm$ 0.91          & 5.27 $\pm$ 1.10          & 4.64 $\pm$ 1.17 \\
 & Visual-haptic  & \textbf{25.45 $\pm$ 7.28} & 3.70 $\pm$ 1.29          & \textbf{4.72 $\pm$ 1.04} & \textbf{5.88 $\pm$ 1.06} & \textbf{4.81 $\pm$ 1.40} \\
\midrule
\multirow{3}{*}{Aerial}
 & Visual-only    & 15.98 $\pm$ 5.17          & \textbf{5.76 $\pm$ 1.43} & 5.09 $\pm$ 1.04          & 4.30 $\pm$ 1.15          & 5.44 $\pm$ 1.71 \\
 & Haptic-only    & 21.52 $\pm$ 5.80          & 4.67 $\pm$ 1.82          & 5.17 $\pm$ 1.24          & 4.95 $\pm$ 1.46          & 5.53 $\pm$ 1.56 \\
 & Visual-haptic  & \textbf{21.52 $\pm$ 8.85} & 5.26 $\pm$ 1.29          & \textbf{5.50 $\pm$ 1.54} & \textbf{5.09 $\pm$ 1.26} & \textbf{5.79 $\pm$ 1.67} \\
\bottomrule
\end{tabular}
\end{table*}

Table~\ref{tab:anova} summarises the repeated-measures ANOVA results for all dependent variables. In the following analyses, F denotes the ANOVA F-statistic with degrees of freedom in parentheses, t denotes the t-statistic for pairwise comparisons, p denotes the significance level, $\eta_p^2$ denotes partial eta-squared as the ANOVA effect size, and g denotes Hedges' g as the pairwise effect size.\

Interface modality had a significant main effect on overall SA, $F(2, 34)$ = 9.35, $p < 0.001$, $\eta_p^2$ = 0.355, meaning that the three interface modalities produced reliably different levels of overall SA with a medium effect size. Scenario also had a significant main effect, $F(1, 17)$ = 6.96, $p = 0.017$, $\eta_p^2$ = 0.291, confirming that the ground scenario produced significantly higher overall SA than the aerial scenario with a large effect. The interaction between modality and scenario was not significant, $F(2, 34)$ = 1.58, $p = 0.227$, $\eta_p^2$ = 0.085, indicating that the modality effect was consistent across both scenarios as shown in Table~\ref{tab:anova}.\

\begin{table*}[t]
\caption{Repeated-measures ANOVA results for all dependent variables, with interface and scenario type as within-subject factors.}
\label{tab:anova}
\centering
\begin{tabular}{lcccccccccccc}
\toprule
 & \multicolumn{4}{c}{\textbf{Interface}} & \multicolumn{4}{c}{\textbf{Scenario}} & \multicolumn{4}{c}{\textbf{Interaction}} \\
\cmidrule(lr){2-5} \cmidrule(lr){6-9} \cmidrule(lr){10-13}
\textbf{Variables} & $F$ & $p$ & $\eta^2_p$ & $1-\beta$ & $F$ & $p$ & $\eta^2_p$ & $1-\beta$ & $F$ & $p$ & $\eta^2_p$ & $1-\beta$ \\
\midrule
Overall SA     & 9.35  & \textbf{$<$.001} & .355 & 1.00 & 6.96  & \textbf{.017}  & .291 & 1.00 & 1.58  & .227 & .085 & .68 \\
Demand         & 3.09  & .072           & .154 & .94  & 32.90 & \textbf{$<$.001} & .659 & 1.00 & 1.50  & .241 & .081 & .69 \\
Supply         & 1.88  & .168           & .100 & .85  & 7.62  & \textbf{.013}  & .309 & 1.00 & 0.19  & .796 & .011 & .13 \\
Understanding  & 7.07  & \textbf{.006}  & .294 & 1.00 & 6.11  & \textbf{.024}  & .264 & 1.00 & 1.18  & .313 & .065 & .58 \\
Workload       & 0.73  & .460           & .041 & .39  & 25.24 & \textbf{$<$.001} & .598 & 1.00 & 0.79  & .450 & .044 & .44 \\
\bottomrule
\end{tabular}

\vspace{4pt}
\footnotesize
\textit{Note.} Interface $df = (2, 34)$. Scenario $df = (1, 17)$. Interaction $df = (2, 34)$. $1-\beta$ = post-hoc achieved power. Greenhouse--Geisser correction applied where sphericity was violated. Bold $p$ values indicate significance at $\alpha = 0.05$.
\end{table*}

As reported in Table~\ref{tab:anova}, post-hoc achieved power was computed for all effects using the observed effect sizes, sample size of 18, and $\alpha$ = 0.05. For the modality factor, power exceeded 0.99 for Overall SA and Understanding, the two variables that reached significance, confirming that the sample was sufficient to detect these effects. Power for the modality main effect on Demand reached 0.94 and on Supply reached 0.85, while Workload remained below 0.40. For the scenario factor, power exceeded 0.99 for all five dependent variables, reflecting the large effect sizes associated with the platform difference. For the modality-by-scenario interaction, power was below 0.80 for all five dependent variables, ranging from 0.13 for Supply to 0.69 for Demand. The interaction effects also showed the smallest partial eta-squared values among the three sources, ranging from $\eta_p^2$ = 0.011 to 0.085.\

\begin{figure}
    \centering
    \includegraphics[width=\linewidth]{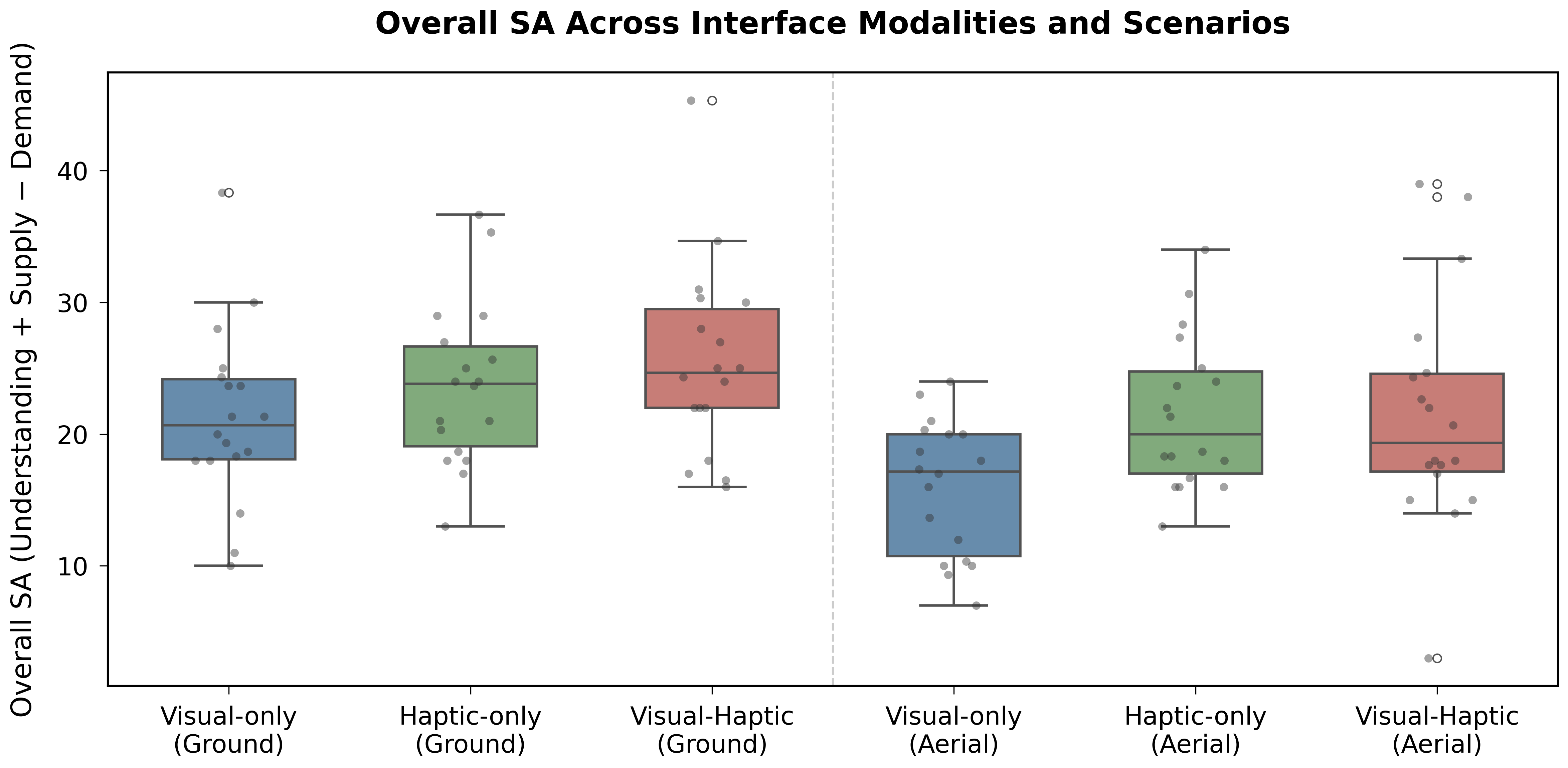}
    \caption{Overall SA scores across all six conditions. Blue, green, and red boxes correspond to the visual-only, haptic-only, and visual-haptic conditions respectively, shown separately for the ground and aerial scenarios. Boxes indicate the interquartile range with median lines, whiskers extend to 1.5 times the interquartile range, and individual points denote participant-level scores.}
    \label{fig:oveallsa}
\end{figure}

\looseness=-1
Fig.~\ref{fig:oveallsa} displays the distributions of overall SA scores across all six conditions. Overall, the ground scenario produced higher SA scores than the aerial scenario across all three modalities, and the modality differences were more visible in the aerial scenario than in the ground scenario. In the ground scenario, medians ranged from 20.67 under visual-only to 23.83 under haptic-only and 24.67 under visual-haptic, with the interquartile ranges largely overlapping. Visual-haptic showed a wider spread extending to higher values, while visual-only had the lowest median among the three conditions. In the aerial scenario, overall SA was lower across all modalities. Visual-only produced the lowest median among all six conditions at 17.17. By contrast, haptic-only and visual-haptic maintained higher medians at 20.00 and 19.33 respectively. The gap between visual-only and the two haptic conditions was visibly larger in the aerial scenario than in the ground scenario.\

\subsection{SA Components}
Among the three SA components, Understanding was the only one showing a significant modality main effect, $F(2, 34)$ = 7.07, $p = 0.006$, $\eta_p^2$ = 0.294, meaning that interface modality reliably influenced how well participants comprehended the situation, with a medium-to-large effect. Neither Demand, $F(2, 34)$ = 3.09, $p = 0.072$, nor Supply, $F(2, 34)$ = 1.88, $p = 0.168$, varied significantly across modalities, indicating that the three interfaces imposed comparable situational demands and provided comparable attentional resources. All three components showed significant scenario effects as reported in Table~\ref{tab:anova}. Demand exhibited the largest scenario effect, $F(1, 17)$ = 32.90, $p < 0.001$, $\eta_p^2$ = 0.659, confirming that the aerial scenario imposed substantially higher situational demand than the ground scenario. Supply also differed between scenarios, $F(1, 17)$ = 7.62, $p = 0.013$, $\eta_p^2$ = 0.309, and Understanding showed a significant scenario effect, $F(1, 17)$ = 6.11, $p = 0.024$, $\eta_p^2$ = 0.264, with lower comprehension scores in the aerial scenario. No modality-by-scenario interaction reached significance for any component, indicating that the modality effects on SA components were consistent across both platforms.\

\begin{figure}
    \centering
    \includegraphics[width=\linewidth]{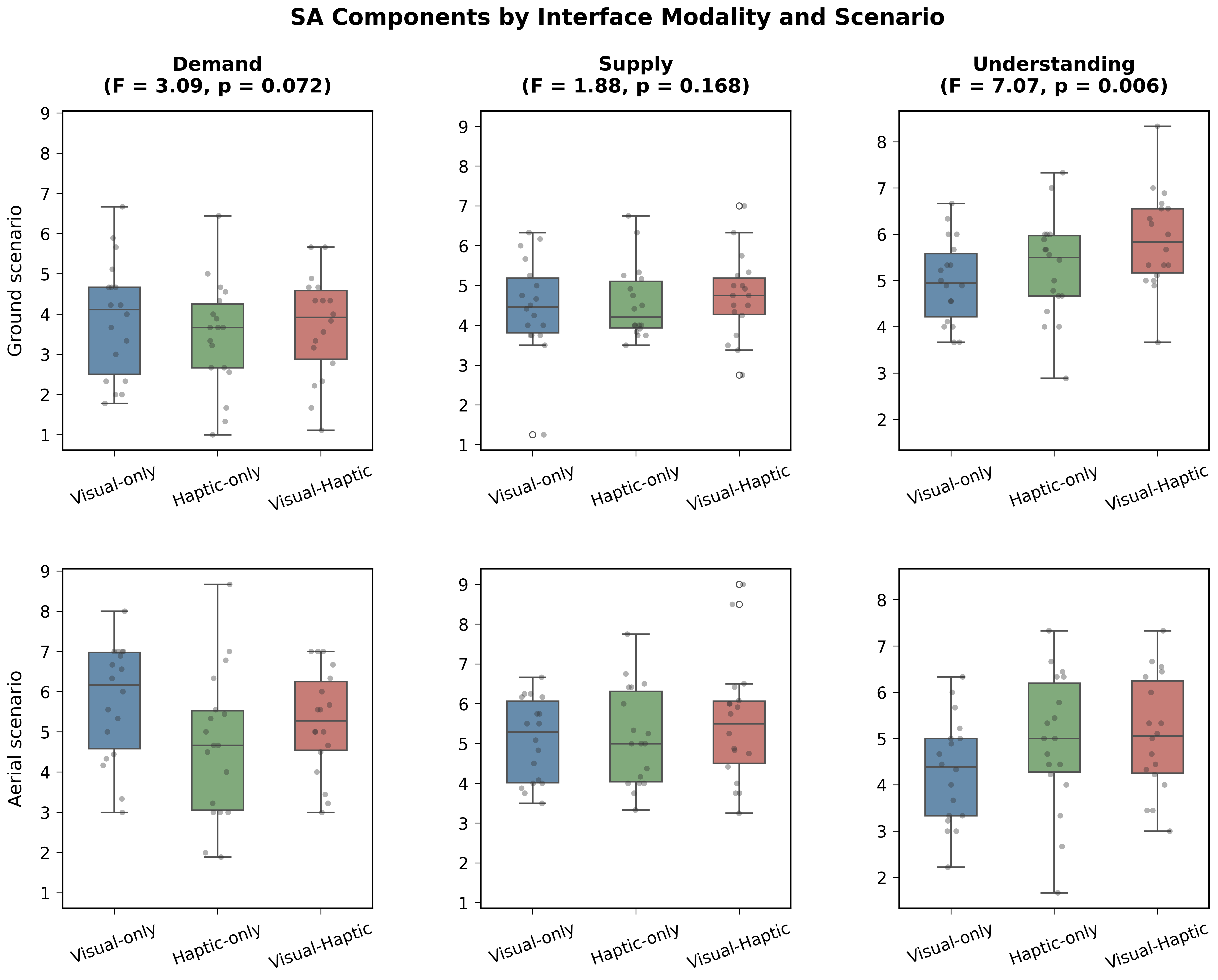}
    \caption{SA component scores by interface modality and scenario. Each column shows one SA component: Demand (left), Supply (centre), and Understanding (right). The top row presents the ground scenario and the bottom row presents the aerial scenario. The modality main effect F statistic and p value from the omnibus ANOVA are displayed above each column. Only Understanding showed a significant modality effect.}
    \label{fig:components}
\end{figure}

Fig.~\ref{fig:components} presents SA component scores by interface modality and scenario. Overall, the most visible modality difference appeared in the Understanding column, whereas Demand and Supply showed minimal separation among the three modalities. Comparing the two rows, the most pronounced cross-scenario shift was in the Demand column. In the Demand column, the three modalities produced largely overlapping distributions in the ground row, with medians at 4.11 for visual-only, 3.67 for haptic-only, and 3.92 for visual-haptic. In the aerial row, all three medians shifted upward to 6.17 for visual-only, 4.67 for haptic-only, and 5.28 for visual-haptic, with visual-only showing the highest median and haptic-only showing the lowest. In the Supply column, the three modalities remained highly overlapping in both rows, with medians ranging from 4.21 to 4.75 in the ground row and from 5.00 to 5.50 in the aerial row. In the Understanding column, a visible ascending pattern from visual-only to haptic-only to visual-haptic was present in both rows. In the ground row, visual-haptic produced a median of 5.83, noticeably higher than visual-only at 4.94. This ascending pattern was maintained in the aerial row, where visual-haptic again showed the highest median at 5.06. Across the two rows, Understanding scores were generally lower in the aerial scenario than in the ground scenario for all three modalities.\

Post-hoc comparisons for Understanding showed that visual-haptic produced significantly higher scores than both visual-only, $t(17)$ = 3.46, $p = 0.009$, $g = 0.86$, a large effect, and haptic-only, $t(17)$ = 2.44, $p = 0.039$, $g = 0.37$, a small-to-medium effect. The comparison between haptic-only and visual-only did not reach significance, $t(17)$ = 1.79, $p = 0.092$, $g = 0.48$, meaning that haptic cues alone did not significantly enhance situational comprehension relative to vision alone. Post-hoc results are reported in Table~\ref{tab:posthoc}.\

\begin{table*}[t]
\caption{Post-hoc pairwise comparisons for dependent variables with significant modality main effects, collapsed across scenario and corrected using the Benjamini--Hochberg procedure.}
\label{tab:posthoc}
\centering
\begin{tabular}{llccc}
\toprule
\textbf{Variables} & \textbf{Comparison} & $t(17)$ & $p_{\text{corr}}$ & \textbf{Hedges' $g$} \\
\midrule
\multirow{3}{*}{Overall SA}
 & Haptic-only vs Visual-only    &  3.27 & \textbf{.007} &  0.77 \\
 & Visual-haptic vs Visual-only  &  3.89 & \textbf{.004} &  0.80 \\
 & Haptic-only vs Visual-haptic  & $-$0.79 & .441          & $-$0.15 \\
\midrule
\multirow{3}{*}{Understanding}
 & Haptic-only vs Visual-only    &  1.79 & .092          &  0.48 \\
 & Visual-haptic vs Visual-only  &  3.46 & \textbf{.009} &  0.86 \\
 & Visual-haptic vs Haptic-only  &  2.44 & \textbf{.039} &  0.37 \\
\bottomrule
\end{tabular}
\end{table*}

\subsection{Perceived Workload and Workload--SA Demand Coupling}
\begin{figure}
    \centering
    \includegraphics[width=\linewidth]{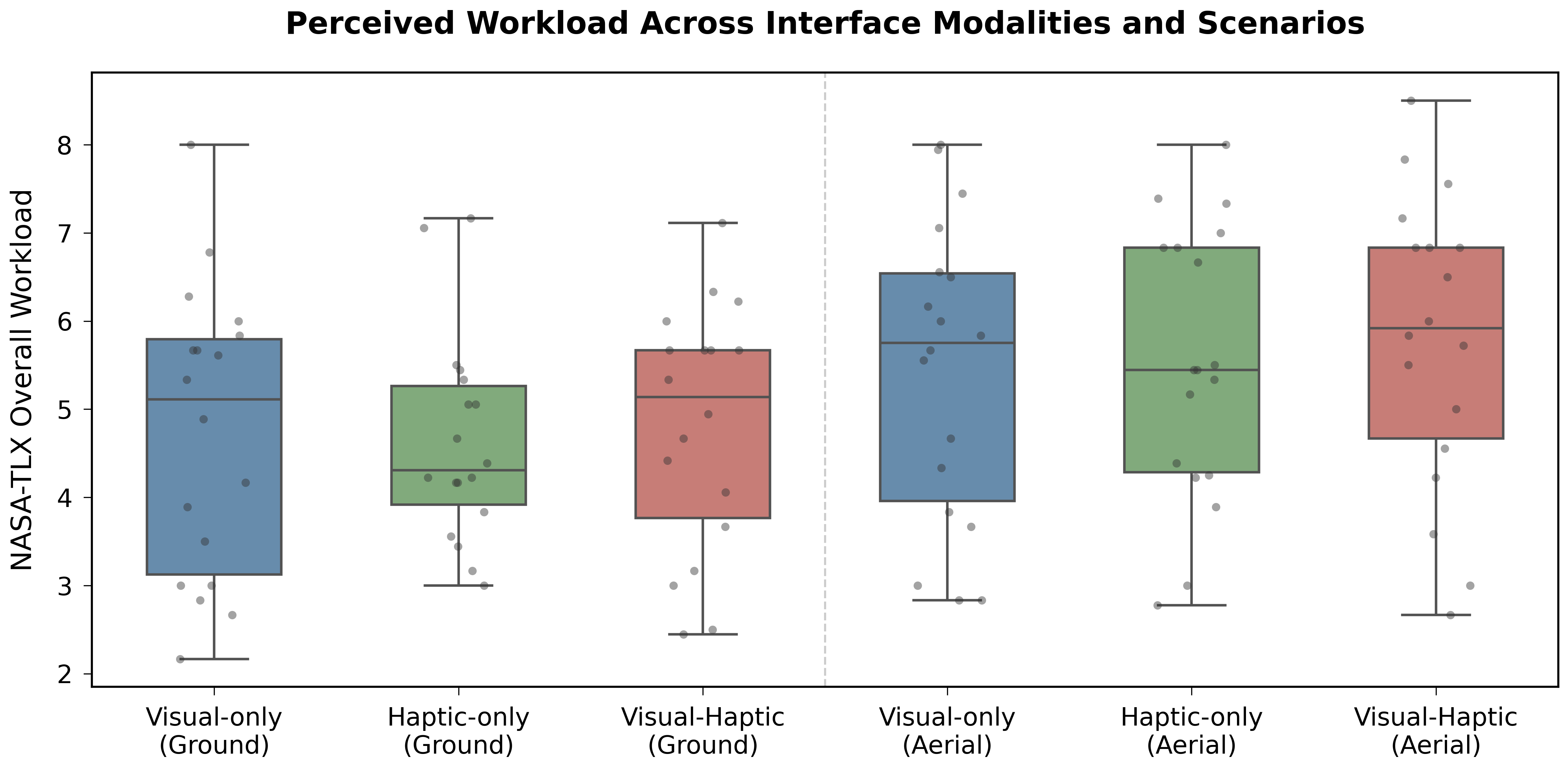}
    \caption{Perceived workload scores across all six conditions. Blue, green, and red boxes correspond to the visual-only, haptic-only, and visual-haptic conditions respectively, shown separately for the ground and aerial scenarios. Boxes indicate the interquartile range with median lines, whiskers extend to 1.5 times the interquartile range, and individual points denote participant-level scores. Workload was consistently higher in the aerial scenario, while no visible modality differences emerged in either scenario.}
    \label{fig:workload}
\end{figure}

Fig.~\ref{fig:workload} displays the distributions of perceived workload across all six conditions. Overall, the three modalities produced similar workload levels within each scenario, while a clear upward shift was visible from the ground to the aerial scenario. In the ground scenario, the three modalities showed overlapping distributions with medians at 5.11 for visual-only, 4.31 for haptic-only, and 5.14 for visual-haptic, and no systematic ordering among the three modalities was apparent. In the aerial scenario, all three medians shifted upward to 5.75 for visual-only, 5.44 for haptic-only, and 5.92 for visual-haptic. The interquartile ranges in the aerial scenario were wider for haptic-only and visual-haptic than for visual-only. Within each scenario, the modality differences remained minimal relative to the scenario difference.\

Interface modality did not significantly affect perceived workload, $F(2, 34)$ = 0.73, $p = 0.460$, $\eta_p^2$ = 0.041, indicating that perceived workload remained comparable across the three interface modalities. The modality-by-scenario interaction was also non-significant, $F(2, 34)$ = 0.79, $p = 0.450$, meaning that this pattern held in both the ground and aerial scenarios. As shown in Table~\ref{tab:descriptives}, mean workload scores in the ground scenario were 4.74 (SD = 1.64) under visual-only, 4.64 (SD = 1.17) under haptic-only, and 4.81 (SD = 1.40) under visual-haptic. In the aerial scenario, workload was higher across all modalities: 5.44 (SD = 1.71) under visual-only, 5.53 (SD = 1.56) under haptic-only, and 5.79 (SD = 1.67) under visual-haptic. A significant scenario main effect confirmed this difference, $F(1, 17)$ = 25.24, $p < 0.001$, $\eta_p^2$ = 0.598, indicating that the aerial scenario imposed substantially higher cognitive demands than the ground scenario regardless of modality.\

\begin{table}[t]
\caption{Spearman rank-order correlations between workload and SA Demand.}
\label{tab:spearman}
\centering
\begin{tabular}{llcc}
\toprule
\textbf{Scenario} & \textbf{Modality} & $\rho$ & $p$ \\
\midrule
\multirow{3}{*}{Ground}
 & Visual-only   & .897 & \textbf{$<$.001} \\
 & Haptic-only   & .569 & \textbf{.014} \\
 & Visual-haptic & .613 & \textbf{.007} \\
\midrule
\multirow{3}{*}{Aerial}
 & Visual-only   & .553 & \textbf{.017} \\
 & Haptic-only   & .595 & \textbf{.009} \\
 & Visual-haptic & .738 & \textbf{$<$.001} \\
\bottomrule
\end{tabular}
\end{table}

\looseness=-1
To examine whether perceived workload and situational demand co-varied across conditions, Spearman rank-order correlations were computed between NASA-TLX workload and SA Demand for each of the six conditions. As reported in Table~\ref{tab:spearman}, the coupling was significant in all six conditions, with $\rho$ ranging from 0.55 to 0.90, confirming a consistent positive association between perceived workload and situational demand. The strongest correlation was observed under visual-only in the ground scenario, $\rho$ = 0.90, $p < 0.001$, suggesting that workload and situational demand were most tightly linked when participants relied solely on visual information in the simpler scenario. The weakest correlation was observed under visual-only in the aerial scenario, $\rho$ = 0.55, $p = 0.017$, though still significant, suggesting that the coupling remained stable even under the higher spatial complexity of the aerial environment.\

\begin{figure}
    \centering
    \includegraphics[width=\linewidth]{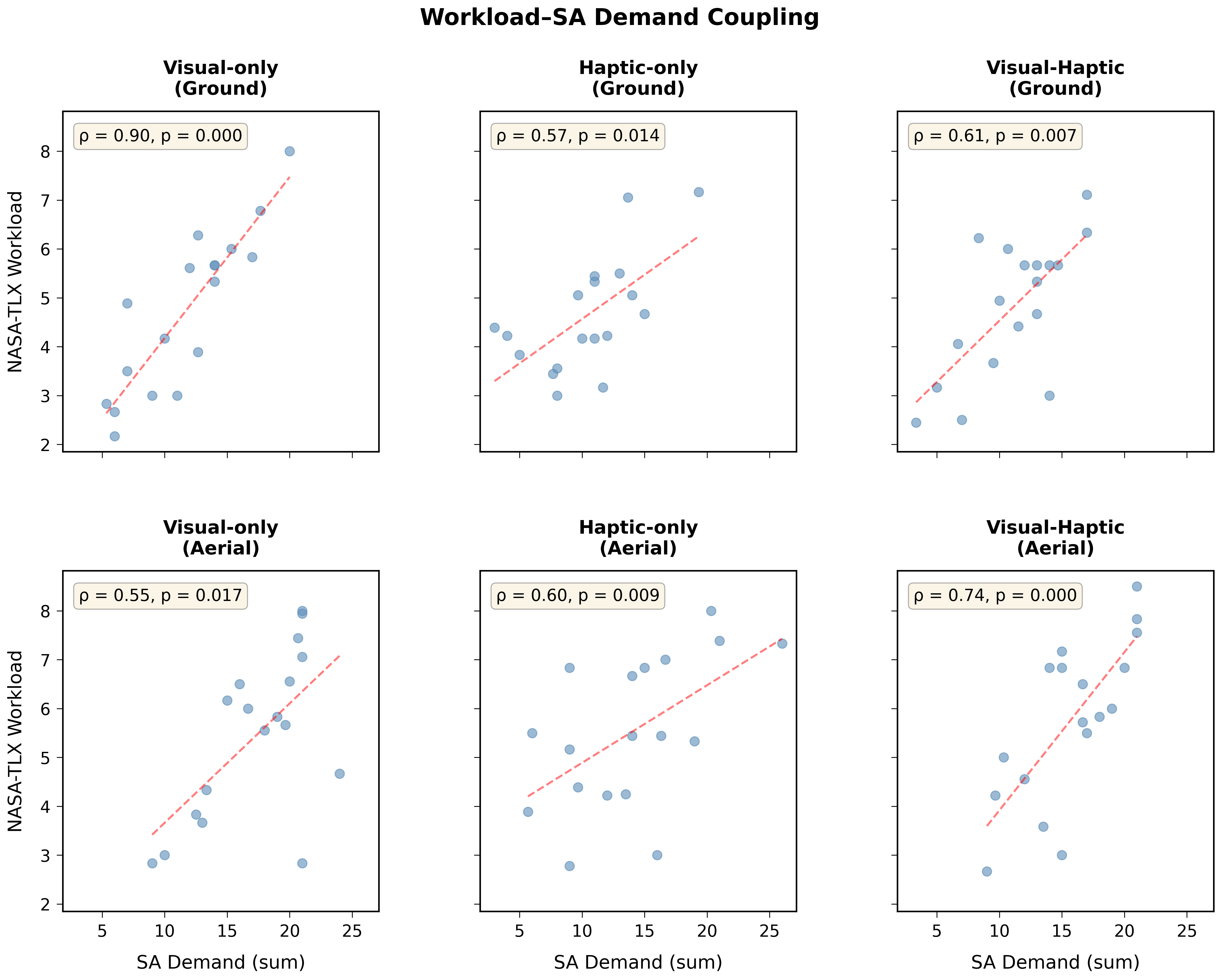}
    \caption{Scatter plots of overall workload against SA Demand for each of the six conditions. Spearman correlation coefficients and p values are shown in each panel. Dashed lines indicate linear trend fits.}
    \label{fig:coupling}
\end{figure}

Fig.~\ref{fig:coupling} presents scatter plots of NASA-TLX workload against SA Demand for each of the six conditions. A positive association between workload and SA Demand was visible in all six panels. The strength and slope varied across conditions. In the ground row, the visual-only panel showed the steepest trend and tightest clustering along the trend line. The haptic-only and visual-haptic ground panels also showed positive trends, though with greater scatter. In the aerial row, the haptic-only and visual-haptic panels maintained clear positive trends with moderate scatter. The visual-only aerial panel showed a weaker but still significant positive trend, with data points more widely scattered around the trend line. The coupling was significant in all six conditions, with $\rho$ ranging from 0.55 to 0.90.\

\section{Discussion and Limitations}
\subsection{Cross-Platform Effects and SA-Workload Coupling}
As reported in Sections 4.1 and 4.2, the aerial scenario produced significantly lower overall SA, higher Demand, and higher workload than the ground scenario, while no modality-by-scenario interaction reached significance. This cross-platform degradation was not uniform across SA components: Demand showed the most pronounced upward shift from ground to aerial, consistent with the largest scenario effect size ($\eta_p^2$ = 0.659); Supply showed only a modest shift ($\eta_p^2$ = 0.309); and Understanding showed a visible downward shift. This pattern indicates that cross-platform SA degradation was primarily driven by elevated situational demand and reduced comprehension rather than diminished processing capacity, and suggests that haptic information may have partially buffered the degradation imposed by the aerial environment, though the interaction did not reach significance.\

The coupling between workload and SA Demand, significant in all six conditions, supports the shared attentional resource mechanism proposed by Wickens et al. \cite{wickens2021engineering}: modality altered SA without shifting the overall workload envelope. Although the visual-only aerial condition produced the weakest coupling, the correlation remained significant, suggesting that the mechanism persisted under higher spatial complexity.\

The aerial scenario differed from the ground scenario in two respects not fully separable in the present design: the aerial environment involved 14 hazard positions across three altitude layers compared with eight on a single plane, and the haptic interface encoded azimuth and elevation on separate vest panels. Transforming vibrotactile cues from a skin-based reference frame into an external spatial representation is known to increase processing demand \cite{heed2015tactile}, and recognition accuracy for torso-based haptic cues drops when azimuth and elevation information are combined \cite{kim2024human}. This confound should be considered when interpreting cross-platform differences.\

\subsection{ Modality Effects on SA and Workload}
As reported in Section 4.3, perceived workload did not vary across modalities, meaning the SA gains were achieved without additional cognitive cost. This is consistent with evidence that well-designed interfaces improve SA without elevating workload \cite{vandemerwe2024agent} and with Multiple Resource Theory \cite{wickens2008multiple, wickens2024misconceptions}, which predicts that distributing information across channels reduces within-channel competition. The present findings extend both accounts to a haptic-only condition not previously tested.\

As reported in Section 4.1, haptic-only and visual-haptic both produced significantly higher overall SA than visual-only, with the difference more pronounced in the aerial scenario. The haptic channel alone therefore appears sufficient to produce the SA gain observed in the combined condition. Previous work comparing only visual-only and visual-haptic could not determine whether this gain originated from the haptic channel itself or multimodal redundancy \cite{wang2025cross}; the present haptic-only condition resolves this ambiguity.\

The component-level analysis in Section 4.2 revealed that this modality effect was localised in Understanding, where only visual-haptic significantly exceeded visual-only. When both channels were available, participants could cross-reference spatial cues from both sources, supporting a more integrated environmental representation. Haptic-only produced equivalent overall SA gains without significantly improving Understanding, suggesting small concurrent shifts across all three components rather than a single dominant mechanism. One possibility consistent with Multiple Resource Theory \cite{wickens2008multiple} is that removing visual scanning demands freed attentional capacity redistributed across SA dimensions. The two conditions thus enhanced SA through distinct cognitive routes, a finding visible only through component-level decomposition. These results identify Understanding as the primary SA component through which interface design exerts its effect.\

\subsection{Limitation and Future Work}
The study was underpowered to detect modality-by-scenario interactions, with post-hoc power ranging from 0.13 to 0.69, and no a priori power analysis was conducted. Future studies should determine sample sizes based on the effect sizes observed here. The ground and aerial scenarios also differed in hazard position count and haptic mapping structure, introducing potential confounds between scenario difficulty and interface complexity. The haptic familiarisation procedure relied on verbal confirmation rather than a formal accuracy criterion, which may have been consequential in the aerial condition; future work could incorporate forced-choice accuracy testing.\

\looseness=-1
SA was measured using post-trial self-report ratings that cannot capture moment-to-moment changes; future work could complement SART with real-time indicators such as eye tracking \cite{zhou2022eye}. Physiological measures including heart rate variability and skin conductance have demonstrated sensitivity to different temporal dynamics of risk perception in automated driving \cite{perellomarch2022physiological}, and electroencephalography could further capture the neural mechanisms underlying the SA--workload coupling observed here. Finally, whether the observed effects generalise to lower-tempo monitoring tasks, longer sessions, or field-representative settings remains to be examined, and individual differences in haptic perception performance warrant further investigation.\


\section{Conclusion}
This study investigated the relationship between human SA and cognitive workload during supervisory monitoring of ground and aerial autonomous vehicles under hazard conditions, examining both cross-platform differences and the effects of interface modality. Three principal findings are: first, aerial monitoring produced significantly lower SA and higher workload than ground monitoring. Component-level analysis revealed that these cross-platform differences were driven by elevated situational Demand and reduced Understanding rather than diminished Supply, highlighting the need for platform-specific interface evaluation. Second, despite these platform differences, perceived workload was consistently coupled with situational demand across all conditions, supporting a stable SA--workload mechanism that persists across heterogeneous platform types. Third, both haptic-only and visual-haptic produced significantly higher overall SA than visual-only without increasing workload, and the two did not differ from each other. The two conditions enhanced SA through distinct cognitive routes: visual-haptic improved Understanding specifically, while haptic-only produced distributed gains across components. Vibrotactile body-worn cues can therefore sustain operator SA as effectively as combined visual-haptic presentation without elevating workload, indicating that the haptic channel alone is sufficient for hazard awareness in supervisory monitoring. These findings suggest that: 1) cross-platform differences in SA and workload are driven by component-specific degradation rather than uniform decline, and evaluation of operator SA in heterogeneous autonomous systems should therefore be conducted at the component level rather than relying on aggregate measures. 2) In operational settings where visual attention is constrained or unavailable, vibrotactile body-worn cues offer a standalone channel for maintaining hazard awareness, reducing dependence on visual displays for safety-critical supervisory monitoring.


%


\ifCLASSOPTIONcaptionsoff
  \newpage
\fi



%

\bibliographystyle{IEEEtran}
\bibliography{references}

%





\end{document}